\documentclass[twocolumn,showpacs,preprintnumbers,amsmath,amssymb,floatfix]{revtex4}
\usepackage{graphicx}% Include figure files
\usepackage{afterpage}% help with floats
\usepackage{amsmath}% for help with equations
\usepackage{dcolumn}% Align table columns on decimal point
\usepackage{bm}% bold math
\usepackage{nicefrac}%inline fractions
\usepackage{siunitx}

\graphicspath{{Biezums_img/}{Jauda_img/}{Trans_img/}{Bildes/}{./}{Temp_img/}}

\begin{document}

\preprint{APS/123-QED}

\title{Spatial dynamics of laser-induced fluorescence in an intense laser beam: experiment and theory in alkali metal atoms}

\author{M.~Auzinsh}
\email{Marcis.Auzins@lu.lv}
\author{A.~Berzins}%
\author{R.~Ferber}%
\author{F.~Gahbauer}%
\author{U.~Kalnins}
\affiliation{Laser Centre, The University of Latvia, 19 Rainis Boulevard, LV-1586 Riga, Latvia}

\date{\today}% It is always \today, today,

\begin{abstract}
We have shown that it is possible to model accurately optical phenomena in intense laser fields by taking into account the 
intensity distribution over the laser beam. We developed a theoretical model that 
divided an intense laser beam into concentric regions, each with a Rabi frequency that corresponds to the intensity in that region,
and solved a set of coupled optical Bloch equations for the 
density matrix in each region. Experimentally obtained magneto-optical resonance curves for the 
$F_g=2\longrightarrow F_e=1$ transition of the $D_1$ line of $^{87}$Rb
agreed very well with the theoretical model up to a laser intensity of around 200~mW/cm$^2$ for a transition whose saturation 
intensity is around 4.5~mW/cm$^2$. 
We have studied the spatial dependence of the fluorescence intensity in an intense laser beam experimentally and theoretically. 
An experiment was conducted whereby a broad, intense pump laser excited the 
$F_g=4\longrightarrow F_e=3$ transition of the $D_2$ line of cesium while a weak, narrow probe beam scanned the atoms within the 
pump beam and excited the $D_1$ line of cesium, whose fluorescence was recorded as a function of probe beam position. 
Experimentally obtained spatial profiles of the fluorescence intensity agreed qualitatively with the predictions of the model. 
\end{abstract}
%32.60.+i Zeeman and Stark effects
%32.80.Xx Level crossing and optical pumping
%32.10.Fn Fine and hyperfine structure
\pacs{32.60.+i,32.80.Xx,32.10.Fn}% PACS, the Physics and Astronomy
\maketitle

\section{\label{Intro:level1}Introduction}
Coherent radiation can polarize the angular momentum distribution of an ensemble of atoms in various ways, creating different 
polarization moments, which modifies the way these atoms will interact with radiation. Carefully prepared spin polarized 
atoms can make the absorption highly dependent on frequency 
(electromagnetically induced transparency~\cite{Harris:1990}), causing large values of the dispersion, which, in turn, 
are useful for such interesting effects as slow light~\cite{Hau:1999} and optical information storage~\cite{Liu:2001}. 
Electric and magnetic fields, external or inherent in the radiation fields, may also influence the time evolution of 
the spin polarization and cause measurable changes in absorption or fluorescence intensity and/or polarization. 
These effects are the basis of many magnetometry schemes~\cite{Scully:1992,Budker:2007}, 
and must be taken into account in atomic clocks~\cite{Knappe:2005} and when searching for 
fundamental symmetry violations~\cite{Budker:2002} or exotic physics such as 
an electric dipole moment of the electron~\cite{Regan:2002}. 
%Polarization created in the excited state can be transferred to 
%the ground state by optical pumping~\cite{Auzinsh:OptPolAt} when the Zeeman sublevels are degenerate at zero magnetic field. 
Sufficiently strong laser radiation creates atomic polarization in excited as well as in the ground state~\cite{Auzinsh:OptPolAt}
The polarization is destroyed when the Zeeman sublevel degeneracy is removed by a magnetic field. 
Since the ground state has a much longer lifetime, very narrow magneto-optical resonances can be created, which 
are related to the ground-state Hanle effect (see ~\cite{Arimondo:1996} for a review). Such resonances were first 
observed in cadmium in 1964~\cite{Lehmann:1964}. 

The theory of magneto-optical resonances has been understood for some time(see \cite{Budker:2002,Alexandrov:2005,Auzinsh:OptPolAt} for a review), 
and bright (opposite sign) resonances have also been observed and explained~\cite{Kazantsev:1984,Renzoni:2001a,Alnis:2001}; the challenge in describing experiments 
lies in choosing the effects to be included in the numerical calculations so as to find a balance between computation time and accuracy.
The optical Bloch equations (OBEs) for the density matrix have been used as early as 1978 to model magneto-optical 
resonances~\cite{Picque:1978}. In order to achieve greater accuracy, later efforts to model signals took into account  
effects such as Doppler broadening, the coherent properties of the laser radiation, 
and the mixing of magnetic sublevels in an external magnetic field to produce more and more 
accurate descriptions of experimental signals~\cite{Auzinsh:2008}. Analytical models can also achieve excellent descriptions of 
experimental signals at low laser powers in the regime of linear excitation where optical pumping plays a negligible 
role~\cite{Castagna:2011,Breschi:2012}. In recent 
years, excellent agreement has been achieved by numerical calculations even when optical pumping plays a role. 
However, as soon as the laser radiation begins to saturate the 
absorption transition, the model's accuracy suffers. The explanation has been that at high radiation intensities, 
it is no longer possible to model the 
relaxation of atoms moving in and out of the beam with a single rate constant~\cite{Auzinsh:1983, Auzinsh:2008}.  
Nevertheless, accurate numerical models of situations in an intense laser field are very desirable, because they 
could arise in a number of experimental situations. Therefore, we have set out to model magneto-optical effects in the presence of 
intense laser radiation by taking better account of the fact that an atom experiences different laser intensity values as it 
passes through a beam. In practice, we solve the rate 
equations for the Zeeman coherences for different regions of the laser beam with a value of the Rabi frequency that more 
closely approximates the real situation in that part of the beam.
To save computing time, stationary solutions to the rate equations for Zeeman sublevels and coherences are sought for each region~\cite{Blushs:2004}. 
With this simplification to take into account the motion of atoms through the beam, 
we could now obtain accurate descriptions of experimental signals up to much higher intensities for reasonable computing times. 
Moreover, the model can be used to study the spatial distribution of the laser induced 
fluorescence within the laser beam. We performed such a study theoretically and experimentally using two overlapping lasers: 
one spatially broad, intense pump laser, and a weaker, tightly focused, spatically narrow probe laser. 
The qualitative agreement between experimental and theoretical fluorescence intensity profiles indicates 
that the model is a useful tool for studying fluorescence dynamics
as well as for modelling magneto-optical signals at high laser intensities. 

%Magneto-optical resonances are related to the ground-state Hanle effect, first observed in cadmium in 1964~\cite{Lehmann:1964}.
%The optical Bloch equations (OBEs) have for the density matrix been used as early as 1978 to model magneto-optical 
%resonances~\cite{Picque:1978}. In order to achieve greater accuracy, later efforts to model signals took into account the 
%effects such as Doppler broadening and the mixing of magnetic sublevels in an external magnetic field to produce more and more 
%accurate descriptions of experimental signals~\cite{Auzinsh:2008}. Analytical models can also achieve excellent descriptions of 
%experimental signals at low laser powers in the regime of linear excitation where optical pumping plays a negligible 
%role~\cite{Castagna:2011,Breschi:2012}. Although the numerical models 
%worked also in the nonlinear regime with optical pumping, they would become less accurate at high Rabi frequencies, 
%because saturation effects within the beam made the assumption of a single Rabi frequency over the beam profile untenable. 
%Therefore, we set out to extend the OBEs to 
%larger Rabi frequencies by taking into account the power distribution over the laser beam profile. We present the results of 
%experimental calculations and use them to describe experimentally obtained signals for a typical nonlinear magneto-optical resonance 
%at high laser power densities. Furthermore, we conducted calculations and experiments to investegate the spatial 
%distribution of the intensity distribution within an intense laser beam. 

\section{\label{Theory:level1}Theory}
\begin{figure}
	\includegraphics[width=0.45\textwidth]{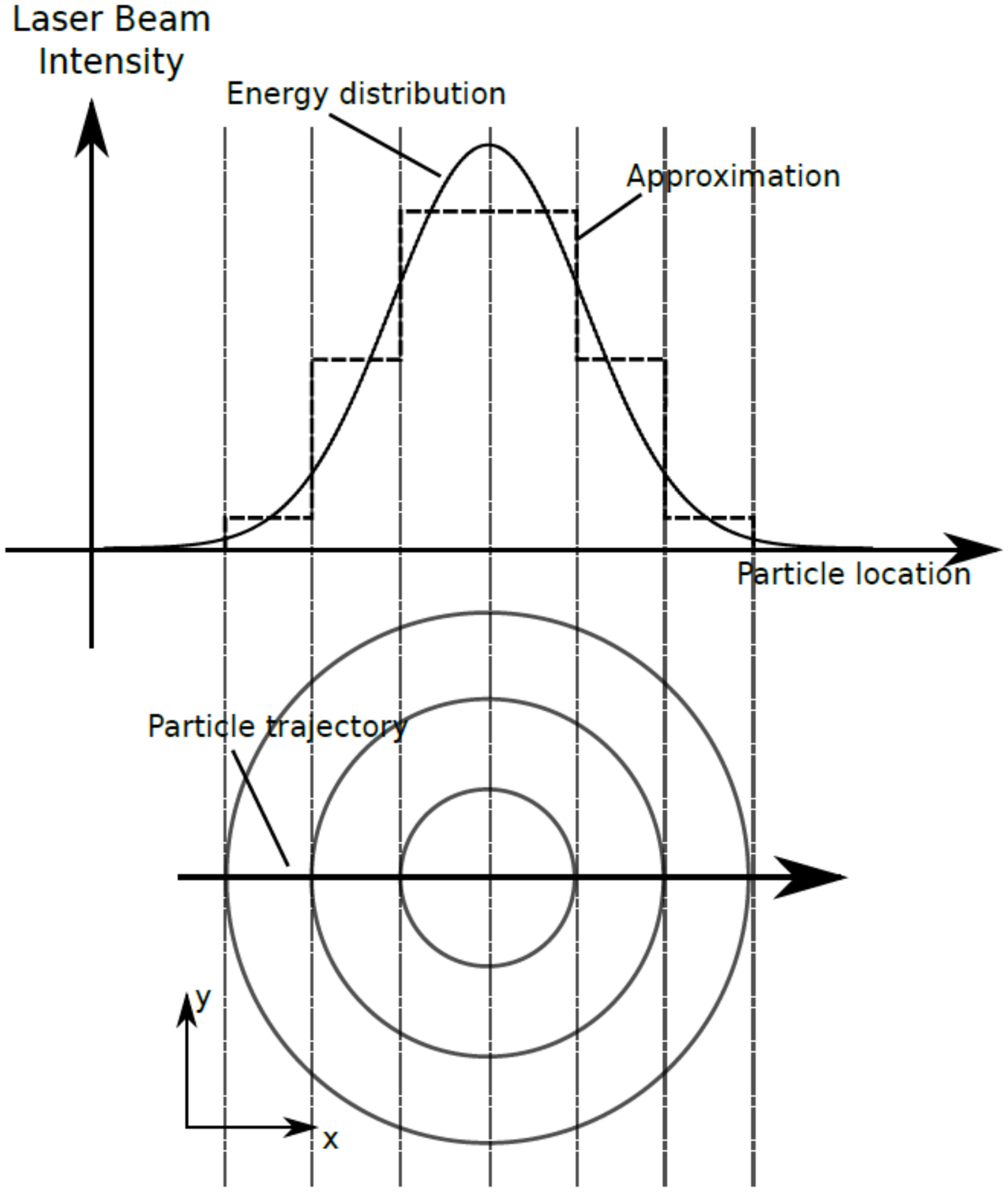}
	\caption{\label{fig:dal1} Laser beam profile split into a number of concentric regions.}
\end{figure}

The theoretical model used here is a further development of previous efforts~\cite{Auzinsh_crossing:2013}, which has been subjected 
to some initial testing in the specialized context of an extremely thin cell~\cite{Auzinsh:2015}.
The description of coherent processes starts with the optical Bloch equation (OBE):
\begin{equation}
i \hbar \frac{\partial \rho}{\partial t} = \left[\hat{H},\rho \right]+ i \hbar \hat{R}\rho,
\end{equation}
where $\rho$ is the density matrix describing the atomic state, $\hat{H}$ is the Hamiltonian of the system, 
and $\hat{R}$ is an operator that describes relaxation.  These equations are transformed into rate equations 
that are solved under stationary conditions in order to obtain the Zeeman coherences in the 
ground ($\rho_{g_ig_j})$ and excited ($\rho_{e_ie_j}$) states\cite{Blushs:2004}. However, when the intensity distribution in the beam is not 
homogeneous, more accurate results can be achieved by dividing the laser beam into concentric regions and solving the 
OBEs for each region separately while accounting for atoms that move into and out of each region as they fly through the 
beam. Figure~\ref{fig:dal1} illustrates the idea. 
The top part of the figure shows the intensity profile of the laser beam, while 
the bottom part of the figure shows a cross-section of the laser beam indicating the concentric regions. 

In order to account for particles that leave one region and enter the next, an extra term must be added to the OBE:
\begin{equation}
\label{eq:transit}
-i \hbar\hat{\gamma_t} \rho+i \hbar\hat{\gamma_t} \rho'.
\end{equation}
In this term, $\rho'$ is the density matrix of the particles entering the region (identical to the density matrix of the 
previous region), and $\hat{\gamma_t}$ is an operator that accounts for transit relaxation. This operator is 
essentially a diagonal matrix with elements $\hat{\gamma}_{t_{ij}}=(\nicefrac{v_{yz}}{s_n})\delta_{ij}$, where $v_{yz}$ 
characterizes the particle speed in the plane perpendicular to the beam and $s_n$ is the linear dimension of the region. 
To simplify matters, we treat particle motion in only one direction and later average with particles that move in the other 
direction. In that case, $\rho'=\rho^{n-1}$. 
Thus, the rate equations for the density matrix $\rho^n$ of the $n$textsuperscript{th} region become 
%\textbf{Can I remove the collision terms with $\hat{\gamma}_c$?}
\begin{align}
\label{eq:rate_region}
i~\hbar \frac{\partial\rho^n}{\partial t} &= \left[ \hat{H},\rho^n\right]+i ~\hbar \hat{R} \rho^n -i~\hbar\hat{\gamma_t}^n\rho^n \notag \\ 
& +i~\hbar\hat{\gamma_t}^n\rho^{n-1}-i~\hbar\hat{\gamma_c}\rho^n+i~\hbar\hat{\gamma_c}\rho^0.
\end{align}
In this equation the relaxation operator $\hat{R}$ describes spontaneous relaxation only and $\hat{\gamma_c}$ is the collisional relaxation 
rate, which, however, becomes significant only at higher gas densities.  

Next, the rotating wave approximation~\cite{Allen:1975} is applied to the OBEs, which yield 
stochastic differential equations that can be simplified by means of the decorrelation 
approach~\cite{Kampen:1976}. Since the measurable quantity is merely light intensity, 
a formal statistical average is performed over the fluctuating phases of these stochastic equations, 
making use of the decorrelation approximation~\cite{Blushs:2004}. As a result, the density matrix 
elements that correspond to optical coherences are eliminated and one is left with rate equations for the 
Zeeman coherences:
\begin{align}
\label{eq:ground}
\frac{\partial \rho_{g_i,g_j}^n}{\partial t} =& \sum_{e_k,e_m}\left(\Xi_{g_ie_m}^n + (\Xi_{e_kg_j}^n)^*\right) d_{g_ie_k}^*d_{e_mg_j}\rho_{e_ke_m}^n \notag \\
& - \sum_{e_k,g_m}(\Xi_{e_kg_j}^n)^*d_{g_ie_k}^*d_{e_kg_m}\rho_{g_mg_j}^n \notag \\ 
& - \sum_{e_k,g_m}\Xi_{g_ie_k}^n d_{g_me_k}^*d_{e_kg_j}\rho_{g_ig_m}^n  \\
& - i\omega_{g_ig_j}\rho_{g_ig_j}^n+\sum_{e_ke_l}\Gamma_{g_ig_j}^{e_ke_l}\rho_{e_ke_l}^n-\gamma_{t}\rho_{g_ig_j}^n \notag \\ 
& + \gamma^{n}_{t}\rho_{g_ig_j}^{n-1}-\gamma^{n}_{c}\rho_{g_ig_j}^n+\gamma_c\rho_{g_ig_j}^0\notag\\
\label{eq:excited}
\frac{\partial \rho_{e_i,e_j}^n}{\partial t} =& \sum_{g_k,g_m}\left((\Xi_{e_ig_m}^n)^* + \Xi_{g_ke_j}^n\right) d_{e_ig_k}^*d_{g_me_j}\rho_{g_kg_m}^n \notag\\
& - \sum_{g_k,e_m}\Xi_{g_ke_j}^nd_{e_ig_k}d_{g_ke_m}^*\rho_{e_me_j}^n \notag \\ 
& - \sum_{g_k,e_m}(\Xi_{e_ig_k}^n)^*d_{e_mg_k}d_{g_ke_j}^*\rho_{e_ie_m}^n \\
& - i\omega_{e_ie_j}\rho_{e_ie_j}^n-\Gamma\rho_{e_ie_j}^n-\gamma^{n}_{t}\rho_{e_ie_j}^n  \notag \\
& +\gamma^{n}_{t}\rho_{e_ie_j}^{n-1}-\gamma_{c}\rho_{e_ie_j}^n \notag.
\end{align}
In both equations, the first term describes population increase and creation of coherence due to induced 
transitions, the second and third terms describe population loss due to induced transitions, the fourth 
term describes the destruction of Zeeman coherences  due to the splitting $\omega{g_ig_j}$, 
respectively, $\omega_{e_ie_j}$ of the Zeeman sublevels in an external magnetic field, 
and the fifth term in Eq.~\ref{eq:excited} describes spontaneous decay with $\Gamma\rho_{e_ie_j}^n$ giving the 
spontaneous rate of decay for the excited state. At the same time the fifth term in Eq.~\ref{eq:ground} 
describes the transfer of population and coherences from the excited state matrix element $\rho_{e_k e_l}$ to 
the ground state density matrix element $\rho_{g_i g_j}$ with rate $\Gamma^{e_k e_l}_{g_i g_j}$. 
These transfer rates are related to the rate of spontaneous decay $\Gamma$ for the excited state. 
Explicit expressions for these $\Gamma^{e_k e_i}_{g_i g_j}$ can be calculated from quantum angular 
momentum theory and are given in~\cite{Auzinsh:OptPolAt}. 
%the fifth term describes spontaneous decay with $\Gamma$ giving the spontaneous rate of decay, 
The remaining terms have been described previously in the context of 
Eqns.~\ref{eq:transit} and~\ref{eq:rate_region}. 
The laser beam interaction is represented by the term
\begin{align}
\Xi_{g_ie_j}= \frac{|\bm\varepsilon^n|^2}{\frac{\Gamma+\Delta\omega}{2}+i \left(\bar{\omega}-\mathbf{k}\cdot \mathbf{v}+\omega_{g_ie_j}\right)},
\end{align} 
where $|\bm\varepsilon^n|^2$ is the laser field's electric field strength in the $n$th region, $\Gamma$ is the spontaneous 
decay rate, $\Delta\omega$ is the laser beam's spectral width, $\bar{\omega}$ is the laser frequency, 
$k\cdot v$ gives the Doppler shift, and $\omega_{g_ie_j}$ is the difference in energy between levels 
$g_i$ and $e_j$. The system of linear equations can be solved for stationary conditions to 
obtain the density matrix $\rho$. 

From the density matrix one can obtain the fluorescence intensity from each region for each velocity group $v$ and given 
polarization $\bm\varepsilon_f$ up to a constant factor of $\tilde{I}_0$~\cite{AuzFerb:OptPolMol, Barrat:1961, Dyakonov:1965}:
\begin{equation}\label{eq:fluorescence}
	I_{n}(v,\bm\varepsilon_f) = \tilde{I}_0\sum\limits_{g_i,e_j,e_k} d_{g_ie_j}^{\ast(ob)}d_{e_kg_i}^{(ob)}\rho_{e_je_k}.
\end{equation}
From these quantities one can calculate the total fluorescence intensity for a given polarization $\bm\varepsilon_f$: 
\begin{align}
	I(\bm\varepsilon_f) = \sum_{n} \sum_{v} f(v)\Delta v \frac{A_{n}}{A} I_{n}(v,\bm\varepsilon_f).
\end{align}
Here the sum over $n$ represents the sum over the different beam regions of relative area 
$\nicefrac{A_{n}}{A}$ as they are traversed by the 
particle, $v$ is the particle velocity along the laser beam, and $f(v)\Delta v$ gives the number of 
atoms with velocity $v\pm \nicefrac{\Delta v}{2}$. 

In practice, we do not measure the electric field strength of the laser field, but the intensity $I=P/A$, where $P$ is the 
laser power and $A$ is the cross-sectional area of the beam. In the theoretical model it is more convenient to use the 
Rabi frequency $\Omega_R$, here defined as follows:
\begin{align}
\label{eq:Rabi}
\Omega_R = k_R \frac{\vert\vert d \vert\vert \cdot \vert\vert \epsilon \vert\vert}{\hbar} \
	 = k_R \frac{\vert\vert d \vert\vert}{\hbar} \sqrt{\frac{2 I}{\epsilon_0 n c}},
\end{align}
where $\vert\vert d \vert\vert$ is the reduced dipole matrix element for the transition in question, $\epsilon_0$ is the 
vacuum permittivity, $n$ is the index of refraction of the medium, $c$ is the speed of light, and $k_R$ is a factor that 
would be unity in an ideal case, but is adjusted to achieve the best fit between theory and experiment since the 
experimental situation will always deviate from the ideal case in some way. 
We assume that the laser beam's intensity distribution follows a Gaussian distribution. We define the average value of 
$\Omega_R$ for the whole beam by taking the weighted average of a Gaussian distribution on the range [0,FWHM/2], where 
FWHM is the full width at half maximum. Thus it follows that the Rabi frequency at the peak of the intensity distribution 
(see Fig.~\ref{fig:dal1}) is $\Omega_R=0.721\Omega_{peak}$. From there the Rabi frequency of each region can be obtained 
by scaling by the value of the Gaussian distribution function.

%\textbf{Probably should add discussion of Rabi frequency similar to what we added in Artur's paper, and also explain how 
%we relate the single region Rabi frequency to the Rabi frequency in each part.}

\section{\label{exp:level1}Experimental setup}
\begin{figure}
	\includegraphics[width=0.45\textwidth]{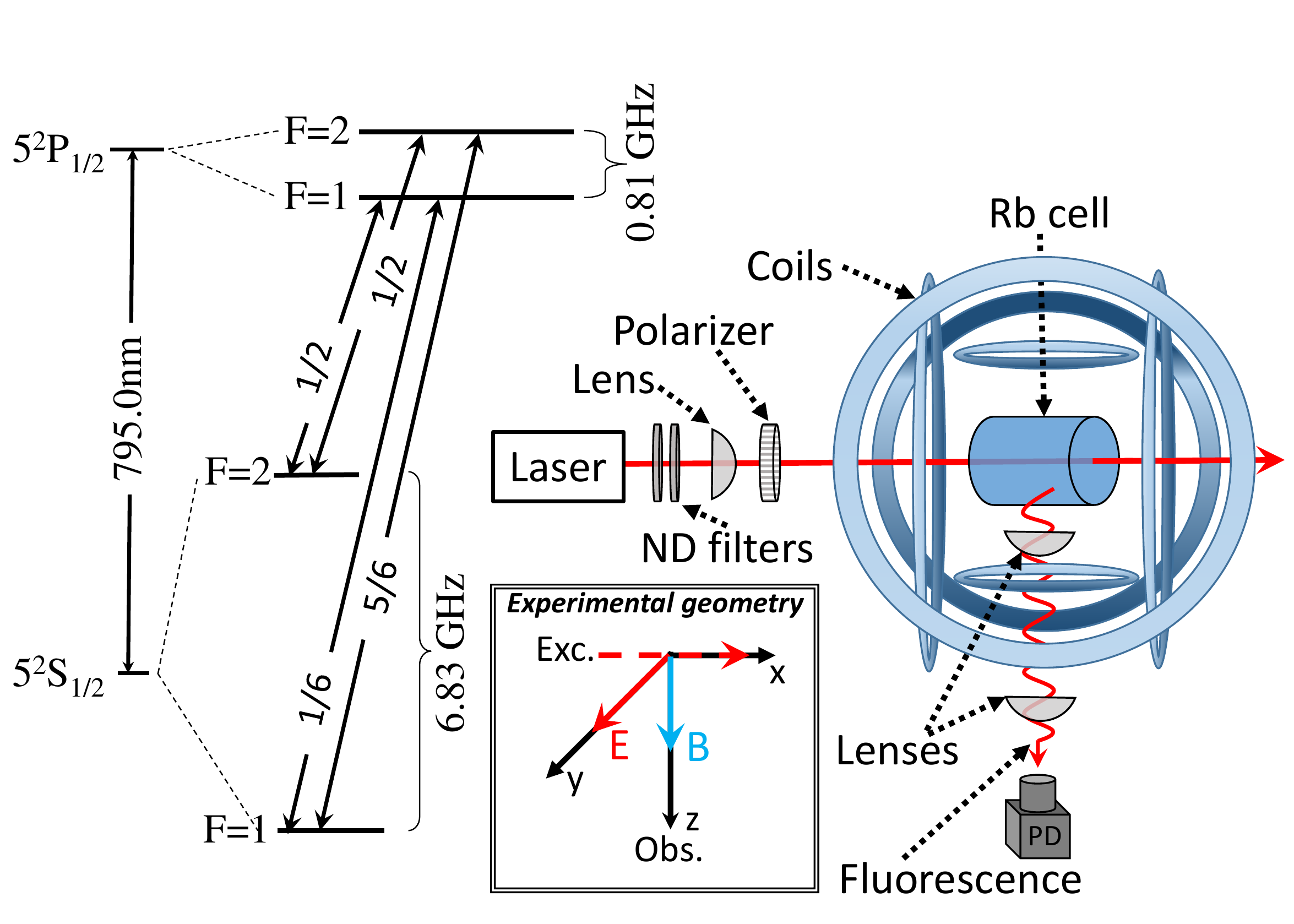}
	\caption{\label{fig:exp_Rb87} (Color online) Basic experimental setup for measuring magneto-optical resonances. The inset 
	on the left shows the level diagram of $^{87}$Rb~\cite{Steck:rubidium87}. The other inset shows the geometrical orientation of the electric 
	field vector \boldmath{E}, the magnetic field vector \boldmath{B}, and laser propagation direction (Exc.) and  
	observation direction (Obs.).}
\end{figure}

%The experimental phenomenon at the heart of this study is known as 
%a magneto-optical resonance, in this case, a dark resonance~\cite{Schmieder:1970,Alzetta:1976}, and it comes about when an 
%external magnetic field breaks the degeneracy of the magnetic sublevels of the ground state, thus destroying the coherences 
%created when these states are degenerate. The effect manifests itself as a reduction in fluorescence at zero magnetic field when 
%atoms in an external magnetic field are excited by linearly 
%polarized light whose polarization vector is perpendicular to the magnetic field.
The theoretical model was tested with two experiments. The first experiment measured magneto-optical resonances on the $D_1$ line of 
$^{87}$Rb and is shown schematically in Fig.~\ref{fig:exp_Rb87}. The experiment has been described elsewhere along with 
comparison to an earlier version of the theoretical model that did not divide the laser beam into separate regions~\cite{Auzinsh:2009}.
%shows the experimental setup used to measure magneto-optical resonances in $^{87}$Rb and test the 
%theoretical model (see~\cite{Auzinsh:2009}).  
The laser was an extended cavity diode laser, whose frequency could be scanned by applying a voltage to a piezo crystal attached to the grating. 
Neutral density (ND) filters were used to regulate the laser intensity, and linear polarization was obtained using a 
Glan-Thomson polarizer. A set of three orthogonal Helmholtz coils scanned the magnetic field along the $z$ axis 
while compensating the ambient field in the other directions. A pyrex cell with a natural isotopic mixture 
of rubidium at room temperature was located at the center of the coils. The total laser induced fluorescence (LIF) 
in a selected direction (without frequency or polarization selection) was detected with 
a photodiode (Thorlabs FDS-100) and data were acquired with a data acquisition card (National Instruments 6024E)
or a digital oscilloscope (Agilent DSO5014). To generate the magnetic field scan with a rate of about 1~Hz, 
a computer-controlled analog signal was applied to a bipolar power supply (Kepco BOP-50-8M). The laser 
frequency was simultaneously scanned at a rate of about 10-20 MHz/s, and it was measured by 
a wavemeter (HighFinnesse WS-7). 
The laser beam was characterized using a beam profiler (Thorlabs BP104-VIS). 

\begin{figure}
	\includegraphics[width=0.45\textwidth]{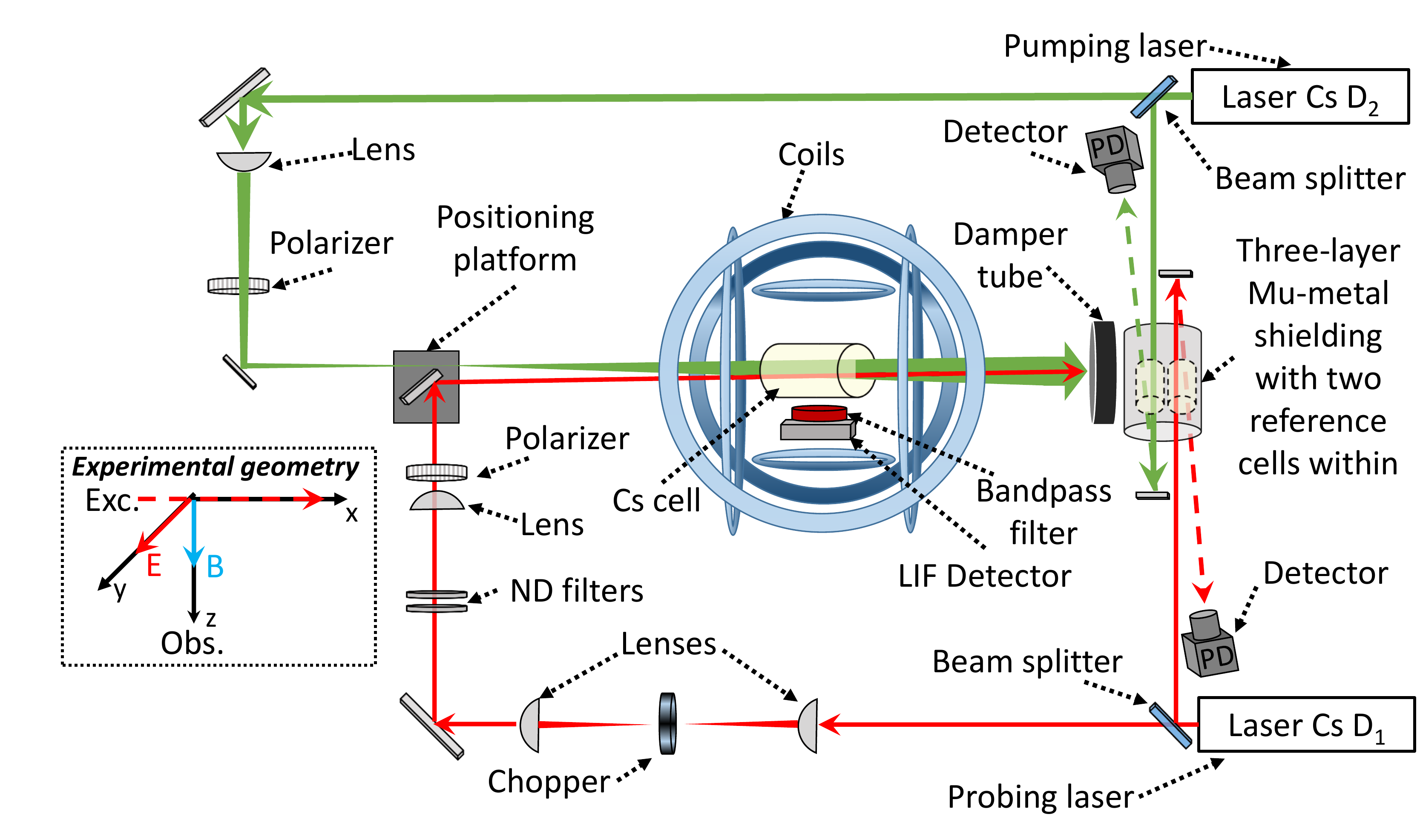}
	\caption{\label{fig:exp_setup} (Color online) Experimental setup for the two-laser experiment. The lasers were stabilized by two Toptica Digilok modules 
	locked to error signals generated from saturated absorption spectroscopy measurements made in a separate, magnetically shielded cell.}
\end{figure}

\begin{figure}
	\includegraphics[width=0.45\textwidth]{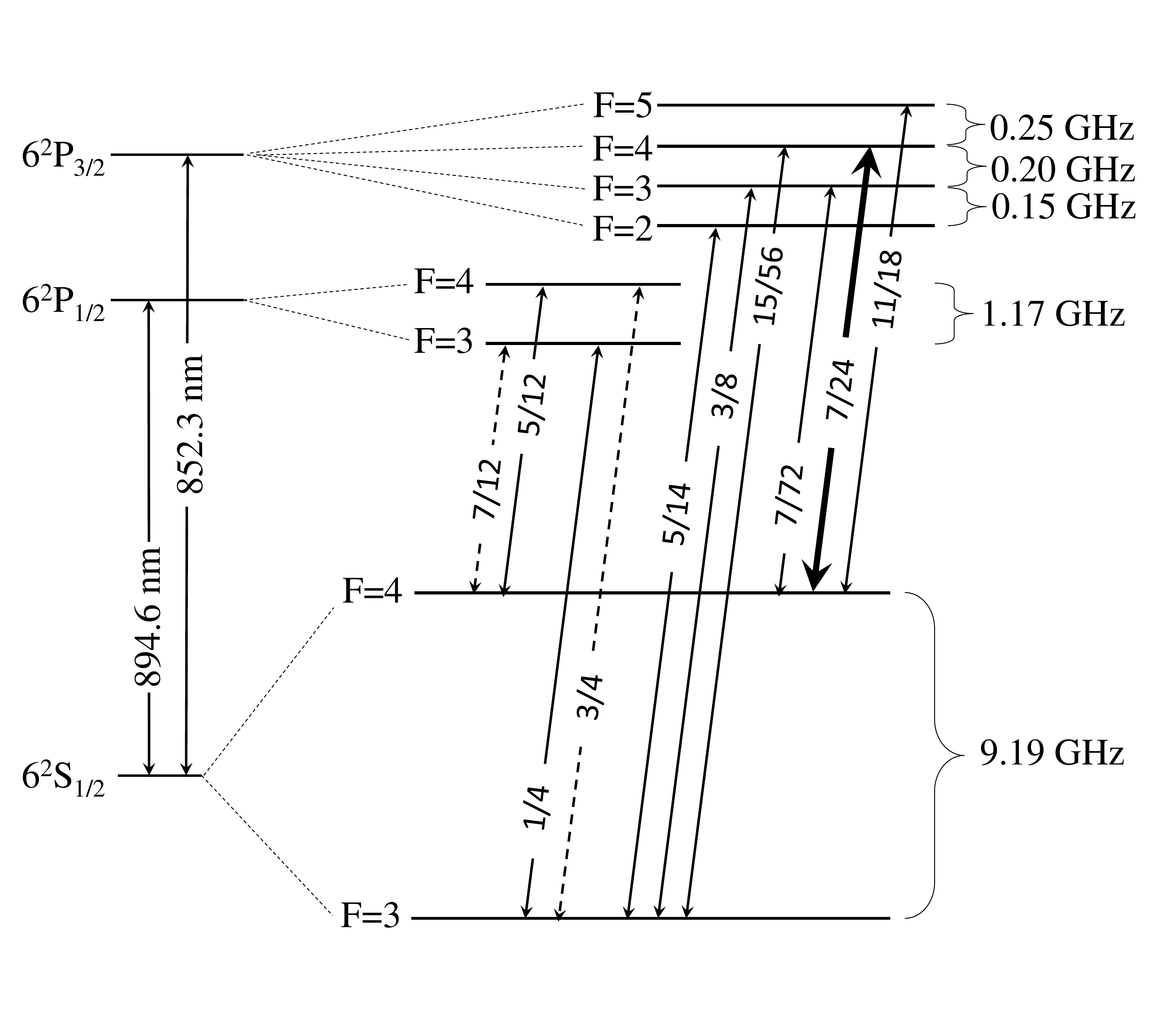}
	\caption{\label{fig:levels} Level scheme for the two-laser experiment. The bold, solid arrow represents the pump laser transition, 
	whereas the arrows with dashed lines represent the scanning laser transitions. Other transitions are given as thin, solid lines.}
\end{figure}

A second experimental setup was used to study the spatial profile of the fluorescence generated by atoms in a 
laser beam at resonance. It is shown in Fig.~\ref{fig:exp_setup}. Here two lasers were used to excite the 
$D_1$ and $D_2$ transitions of cesium. Both lasers were based on distributed feedback diodes from toptica 
(DL100-DFB). One of the lasers (Cs $D_2$) served as a pump laser with a spatially broad and intense beam, 
while the other (Cs $D_1$), 
spatially narrower beam probed the fluorescence dynamics within the pump beam. 
Figure~\ref{fig:levels} shows the level scheme of the excited transitions. 
Both lasers were stabilized with saturation absorption signals from cells shielded by three layers of mu-metal. 
Mu-metal shields were used to avoid frequency drifts due to the magnetic field scan performed in the experiment and other 
magnetic field fluctuations in the laboratory.

A bandpass filter (890~nm $\pm$ 10~nm) was placed before the photodiode. 
To reduce noise from the intense pump beam, the probe beam was modulated by placing a mechanical 
chopper near its focus, and the fluorescence signal was passed through a lock-in amplifier and recorded 
on a digital oscilloscope (Yokogawa DL-6154).  
The probe laser was scanned through the pump laser beam profile using 
a mirror mounted on a moving platform (Nanomax MAX301 from Thorlabs) with a scan range of 8 mm in one dimension. 
The probing beam itself had a full width at half maximum (FWHM) 
diameter of~\SI{200}{\micro\metre} with typical laser power of~\SI{100}{\micro\watt}. 
The pump beam width was~\SI{1.3}{\milli\metre} (FWHM) and its power was~\SI{40}{\milli\watt}. This laser beam diameter was achieved by letting the 
laser beam slowly diverge after passing the 
focal point of a lens with focal length of~\SI{1}{\metre}. The pump laser beam diverged slowly enough to be effectively 
constant within the vapor cell.
The probe beam was also focussed by the same lens to reach its focus point inside the cell. 

%The vapour cell was a 2.5~cm long Pyrex tube with optical quality windows. Its diameter was 2.5~cm. 
%To minimize scattered light entering the detector the cell was wrapped in black paper with 
%exposed tube ends and a $5\times2$~mm$^2$ opening in one side for fluorescence measurements.

%To detect the fluorescence we used a photodiode (Thorlabs FDS100) mounted behind 
%a bandpass filter (890 $\pm$ 10~nm) to minimize background signal at the desired wavelength. 
%Additionally we used damper tube which prevented scattered light from both laser beams 
%from entering the detector after hitting the damper.

\section{\label{1laser:level1}Application of the model to magneto-optical signals obtained for high laser power densities}
\begin{figure}[htpb]
	\includegraphics[width=0.45\textwidth]{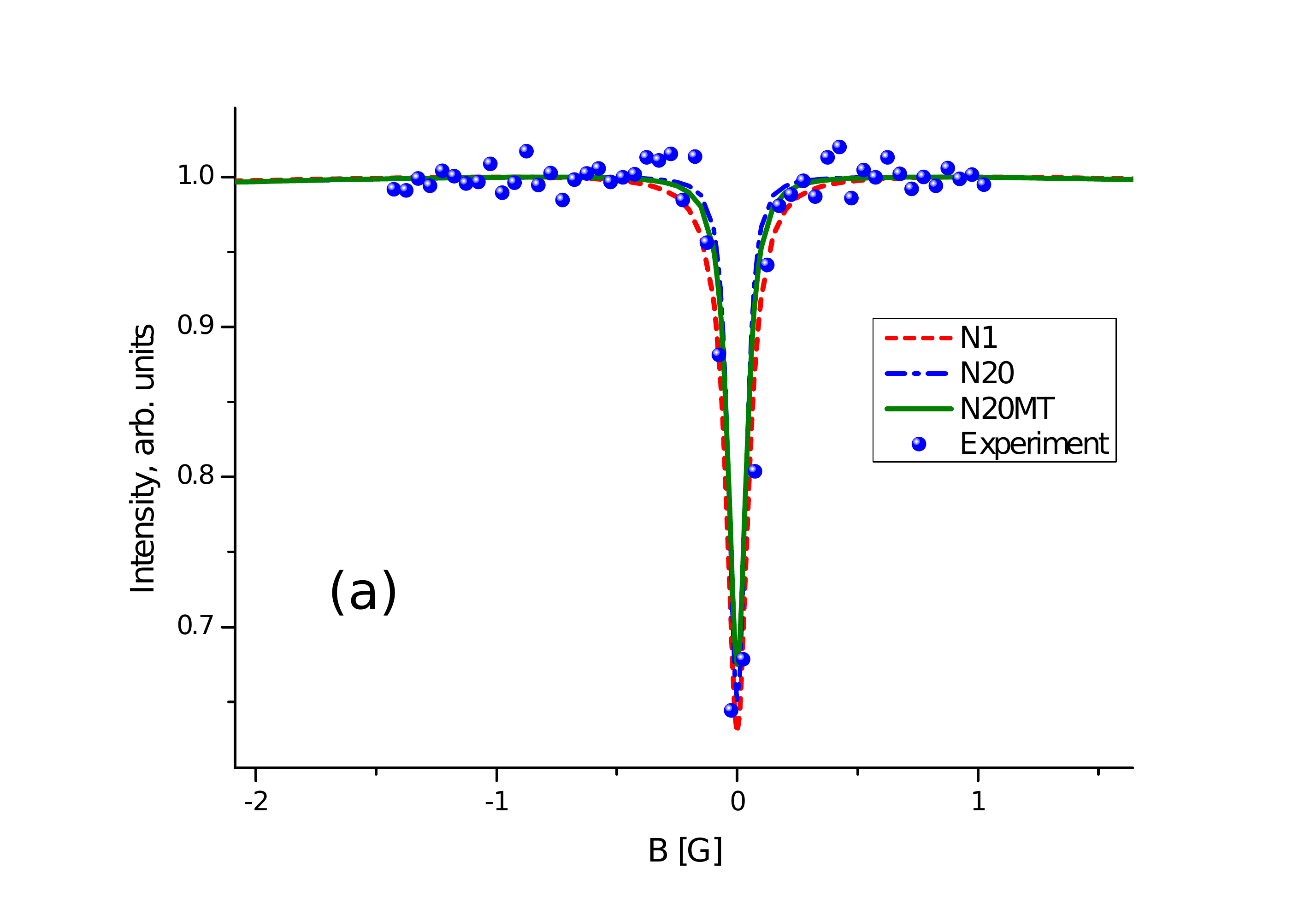}
	\includegraphics[width=0.45\textwidth]{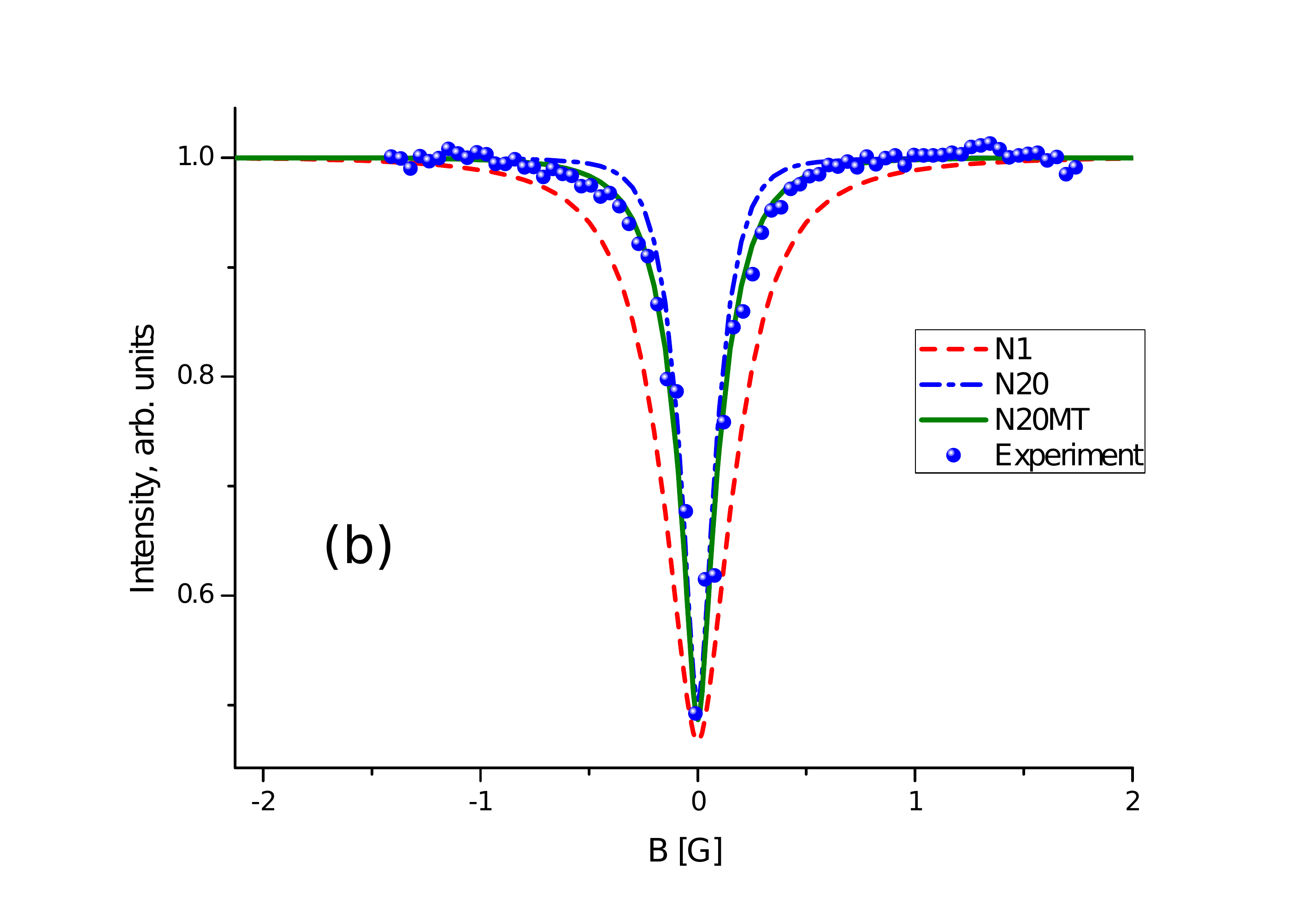}
	\includegraphics[width=0.45\textwidth]{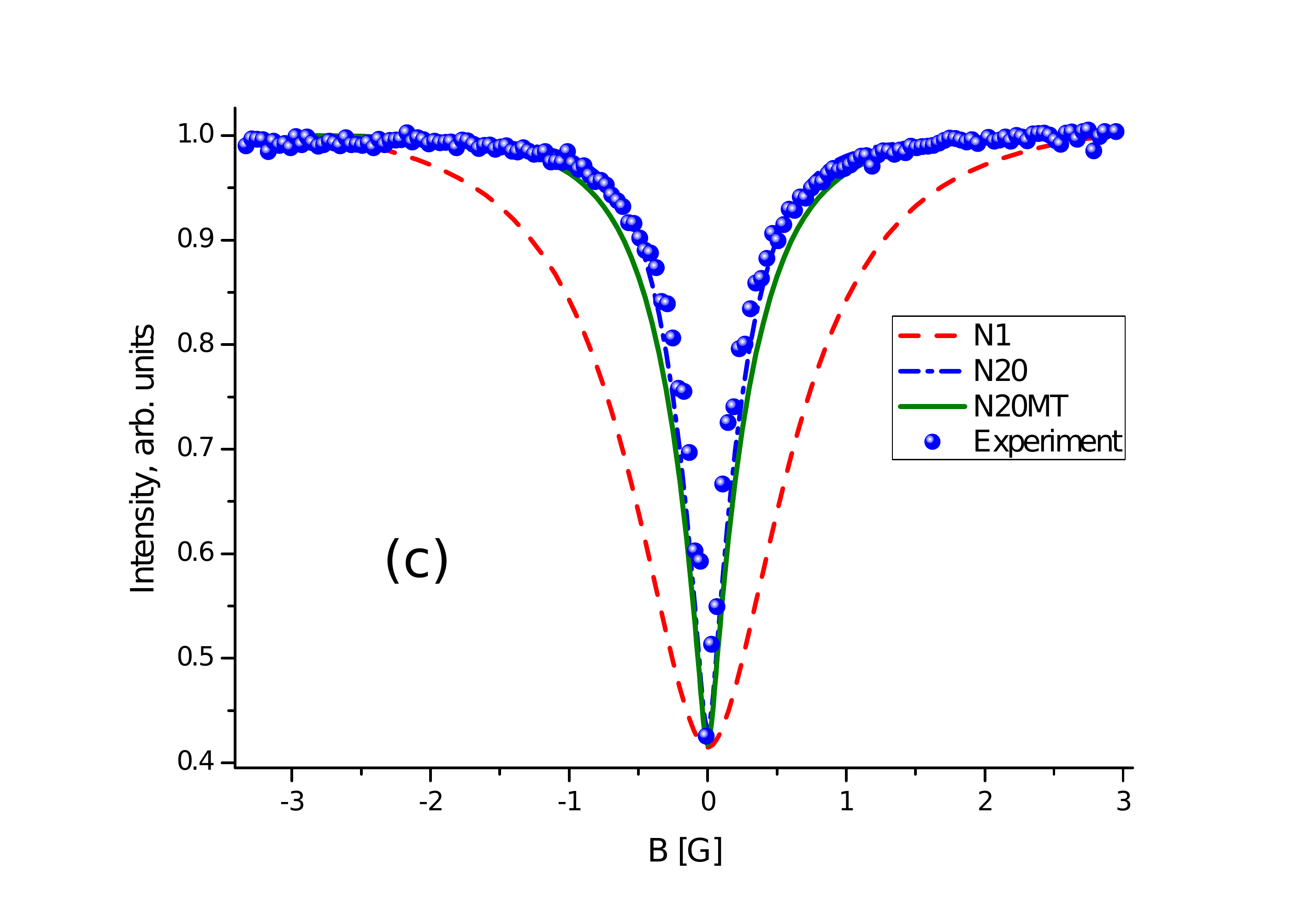}
	\caption{(Color online) Magneto-optical resonances for the $F_g=2\longrightarrow F_e=1$ transition of the $D_1$ line of $^{87}$Rb. 
	Filled circles represent experimental measurements for (a) 28 $\mu$W ($\Omega_R$=2.5 MHz) 
	(b) 280 $\mu$W ($\Omega_R$=8.0 MHz), and 
	(c) 2800  $\mu$W ($\Omega_R$=25 MHz). 
	Curve N1 (dashed) shows the results of a theoretical model that uses one Rabi frequency to model the entire beam profile. 
	Curve N20 (dash-dotted)  shows the result of the calculation when the laser beam profile is divided into 20 concentric circles, 
	and the optical Bloch equations are solved separately for each circle. 
	Curve N20MT (solid) shows the results for a calculation with 20 concentric regions when trajectories are taken into account 
	that do not pass through the center of the beam.  
	}
	\label{fig:one_laser_exp}
\end{figure}

As a first test for the numerical model with multiple regions inside the laser beam, we used the model to calculate the 
shapes of magneto-optical resonances for $^{87}$Rb in an optical cell. The experimental setup was described earlier 
(see Fig.~\ref{fig:exp_Rb87}).   Figure~\ref{fig:one_laser_exp}(a)--(c) 
show experimental signals (markers) and theoretical calculations (curves) of magneto-optical signals in the 
$F_g=2\longrightarrow F_e=1$ transition of the $D_1$ line of $^{87}$Rb. Three theoretical curves are shown: 
curve N1 was calculated assuming a laser beam with a single average 
intensity; curve N20 was calculated using a laser beam divided into 20 concentric regions; curve N20MT was calculated 
in the same way as curve N20, but furthermore the results were averaged also over trajectories that did not pass through 
the center. At the relatively low Rabi frequency of $\Omega_R = 2.5$~MHz [Fig.~\ref{fig:one_laser_exp}(a)] 
all calculated curves practically coincided and described well the experimental signals. The single region model 
treats the beam as a cylindrical beam with an intensity of 2~mW/cm$^2$, which is below the saturation intensity for 
that transition of 4.5~mW/cm$^2$~\cite{Steck:rubidium87}. When the laser intensity was 20~mW/cm$^2$ ($\Omega_R = 8.0$~MHz), 
well above the saturation intensity, model N1 is no longer adequate for describing the experimental signals 
and model N20MT works slightly better [Fig.~\ref{fig:one_laser_exp}(b)]. 
In particular, the resonance becomes sharper and sharper as the intensity increases, and models 
N20 and N20MT reproduce this sharpness. Even at an intensity around 200~mW/cm$^2$ ($\Omega_R = 25$~MHz),  
the models with 20 regions describe the shape of the experimental curve quite well, 
while model N1 describes the experimental results poorly in terms of width and overall shape [Fig.~\ref{fig:one_laser_exp}(c)]. 

\section{\label{distribution:level1}Investigation of the spatial distribution of fluorescence in an intense laser beam}
\subsection{Theoretical investigation of the spatial dynamics of fluorescence in an extended beam}
In order to describe the magneto-optical signals in the previous sections, the fluorescence from all concentric beam regions 
in models N20 and N20MT was summed, since usually experiments measure only total fluorescence (or absorption), especially if 
the beams are narrow. 
However, solving the optical Bloch equations separately for different concentric regions of the laser beam, it is possible 
to calculate the strength of the fluorescence as a function of distance from the center of the beam. With an appropriate 
experimental technique, the distribution of fluorescence within a laser beam could also be measured. 

\begin{figure}[htpb]
	\includegraphics[width=0.45\textwidth]{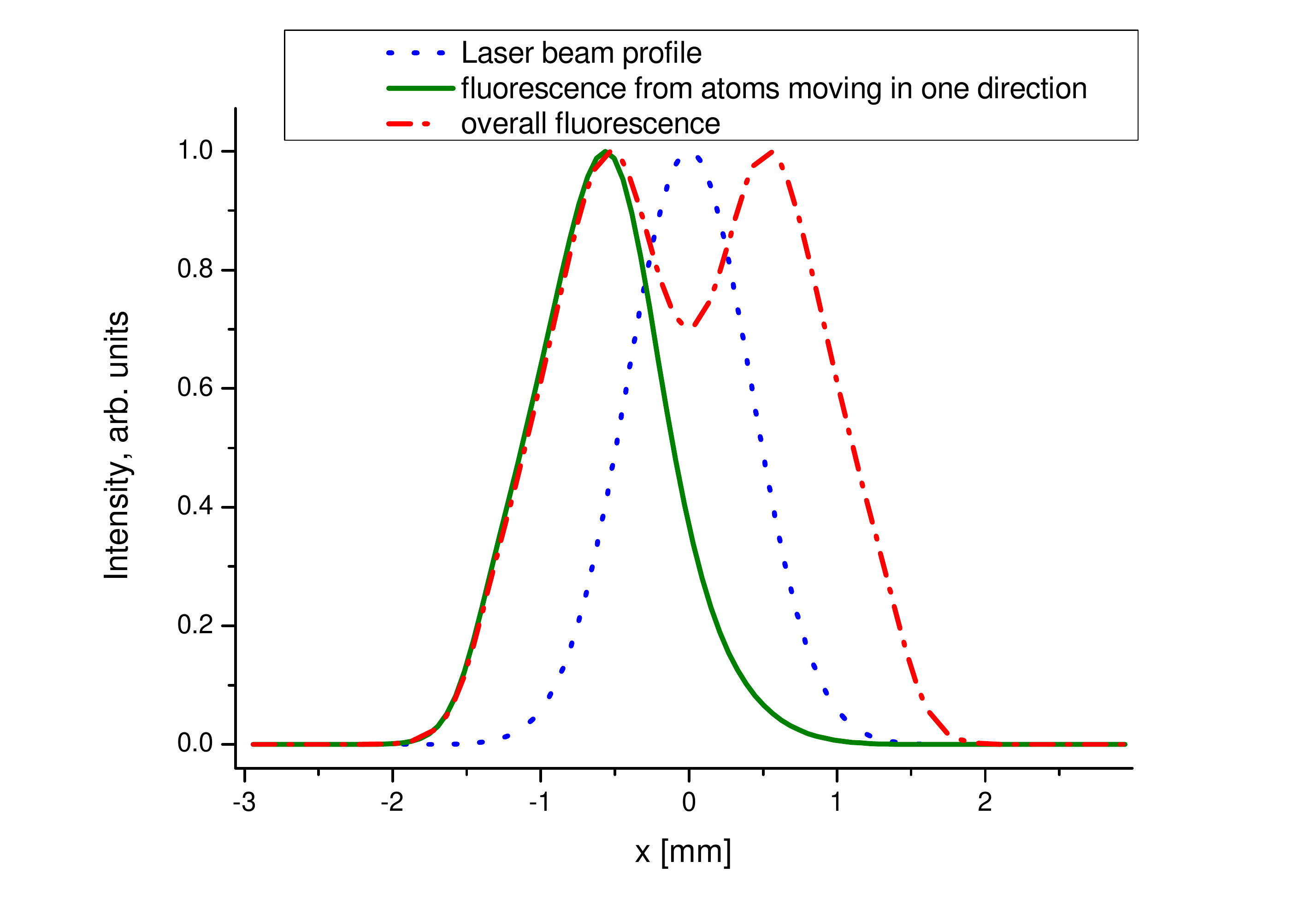}
	\caption{(Color online) Fluorescence distribution as a function of position in the laser beam.
	Dotted (blue) line---laser beam profile, solid (green) line---fluorescence from atoms moving in one direction;
	dash-dotted (red) line---the overall fluorescence as a function of position that results 
	from averaging all beam trajectories. Results from theoretical calculations.}
	\label{fig:dynamics}
\end{figure}

\begin{figure}[htpb]
	\includegraphics[width=0.45\textwidth]{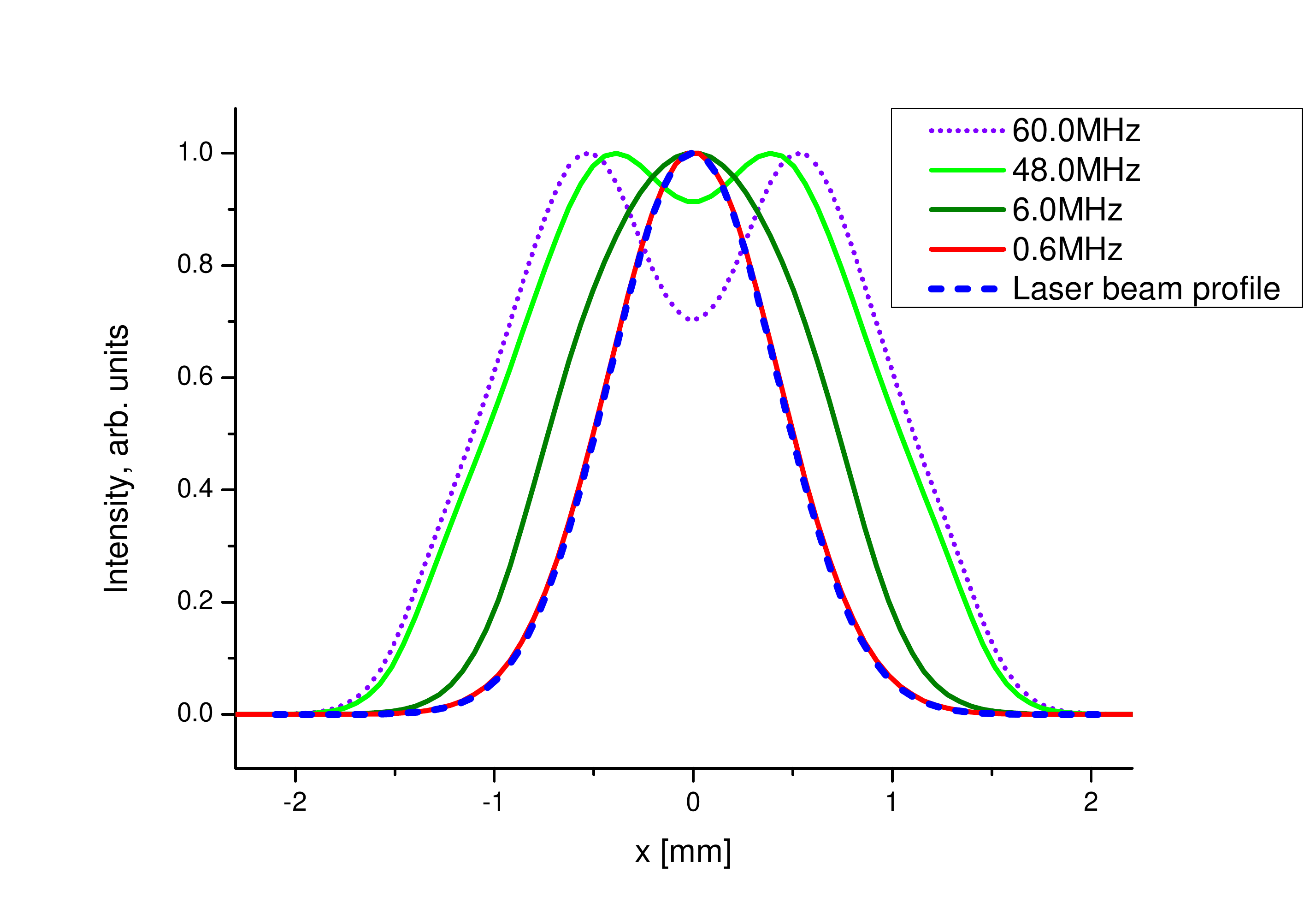}
	\caption{(Color online) Fluorescence distribution as a function of position in the laser beam for various values of the Rabi frequency. 
	Results from theoretical calculations. As the Rabi frequency increases, the distribution becomes broader.}
	\label{fig:dynamics_rabi}
\end{figure}

Figure~\ref{fig:dynamics} shows the calculated fluorescence distribution as a function of position in the laser beam. 
As atoms move through the beam in one direction, the intense laser radiation optically pumps the ground state. In a very 
intense beam, the ground state levels that can absorb light have emptied even before the atoms reach the center 
(solid, green curve). Since atoms are actually traversing the beam from all directions, the result is a fluorescence profile with a 
reduction in intensity near the center of the beam (dashed, red curve). 
The effect of increasing the laser beam intensity (or Rabi frequency) can be seen in Fig.~\ref{fig:dynamics_rabi}.
At a Rabi frequency of $\Omega_R=0.6$ MHz, the fluorescence profile tracks the intensity profile of the laser beam 
exactly. When the Rabi frequency is increased ten times ($\Omega_R=6.0$ MHz), which corresponds to an intensity 
increase of 100, the fluorescence profile already appears somewhat deformed and wider than the actual laser beam profile. 
At Rabi frequencies of $\Omega_R=48.0$ MHz and greater, the fluorescence intensity at the center of the intense laser beam 
is weaker than towards the edges as a result of the ground state being depleted by the intense radiation 
before the atoms reach the center of the laser beam.
 
\subsection{Experimental study of the spatial dynamics of excitation and fluorescence in an intense, extended beam}

\begin{figure*}
	\includegraphics[width=0.45\textwidth]{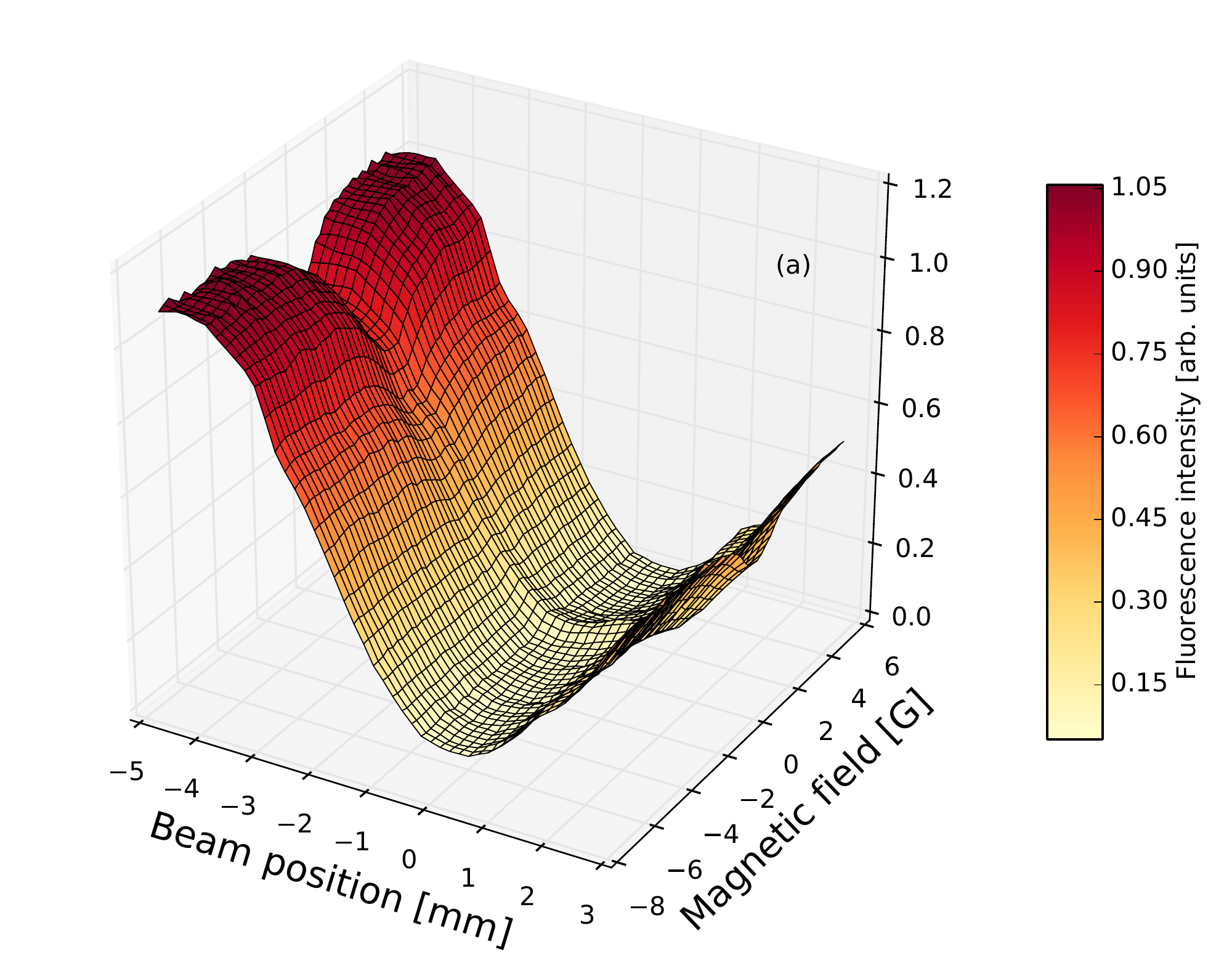}
	\includegraphics[width=0.45\textwidth]{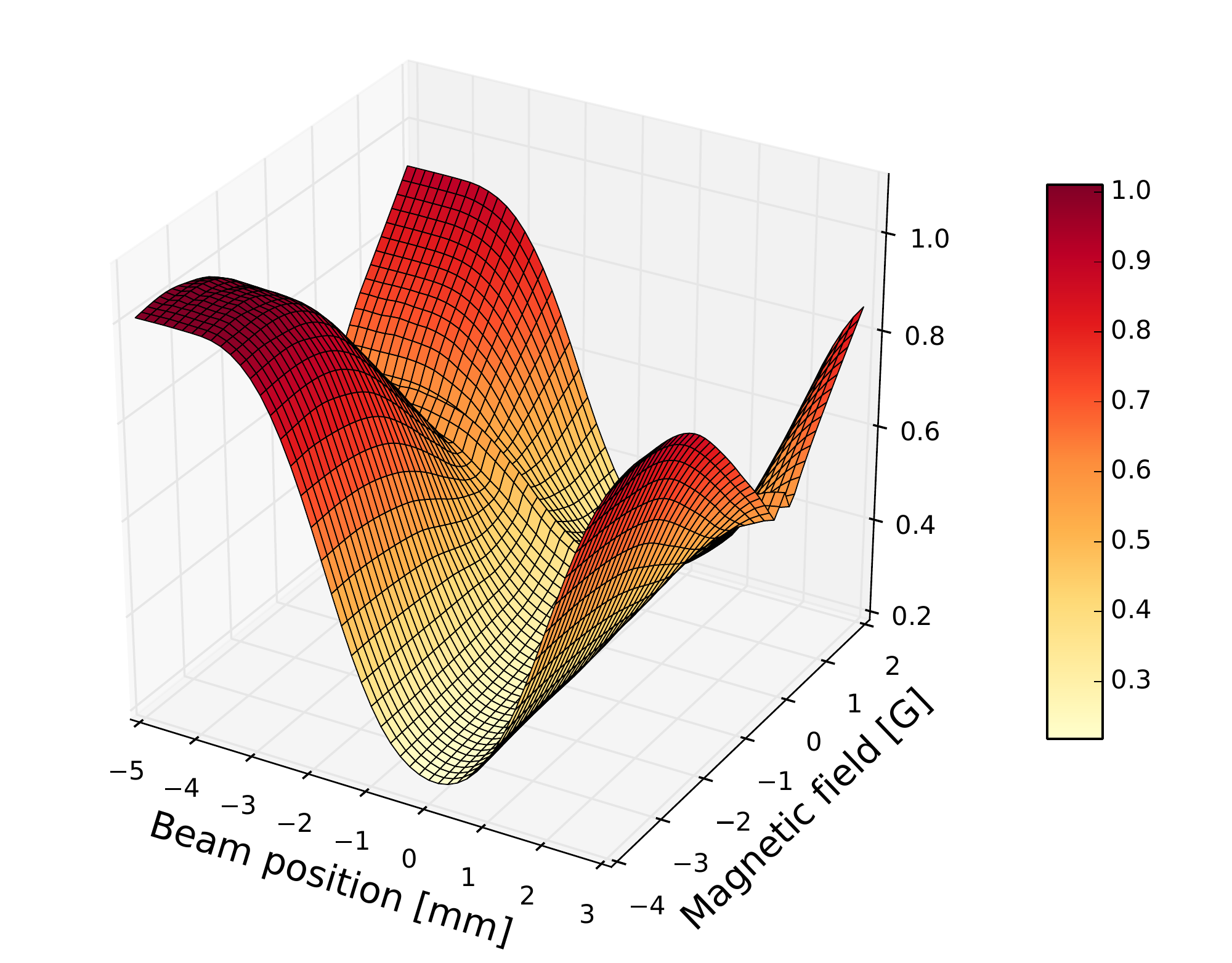}
	\caption{\label{fig:p44_s43} (Color online) Magneto-optical resonances produced for various positions of the probing laser beam 
	($F_g=4\longrightarrow F_e=3$ transition of the $D_1$ line of cesium) with respect to the pump laser beam 
	($F_g=4\longrightarrow F_e=4$ transition of the $D_2$ line of cesium): 
	(a) experimental results and (b) theoretical calculations.  }
\end{figure*}

\begin{figure*}
	\includegraphics[width=0.45\textwidth]{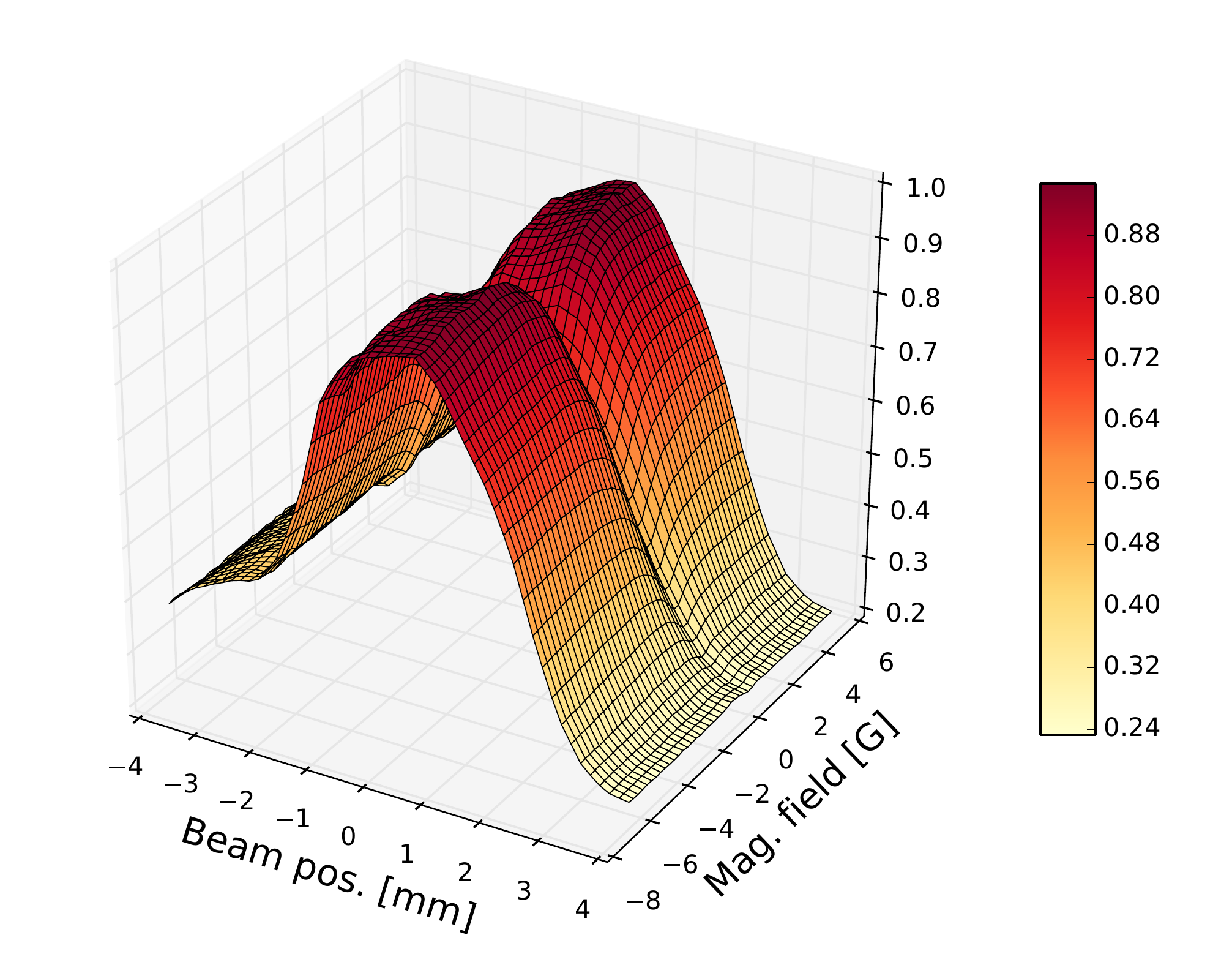}
	\includegraphics[width=0.45\textwidth]{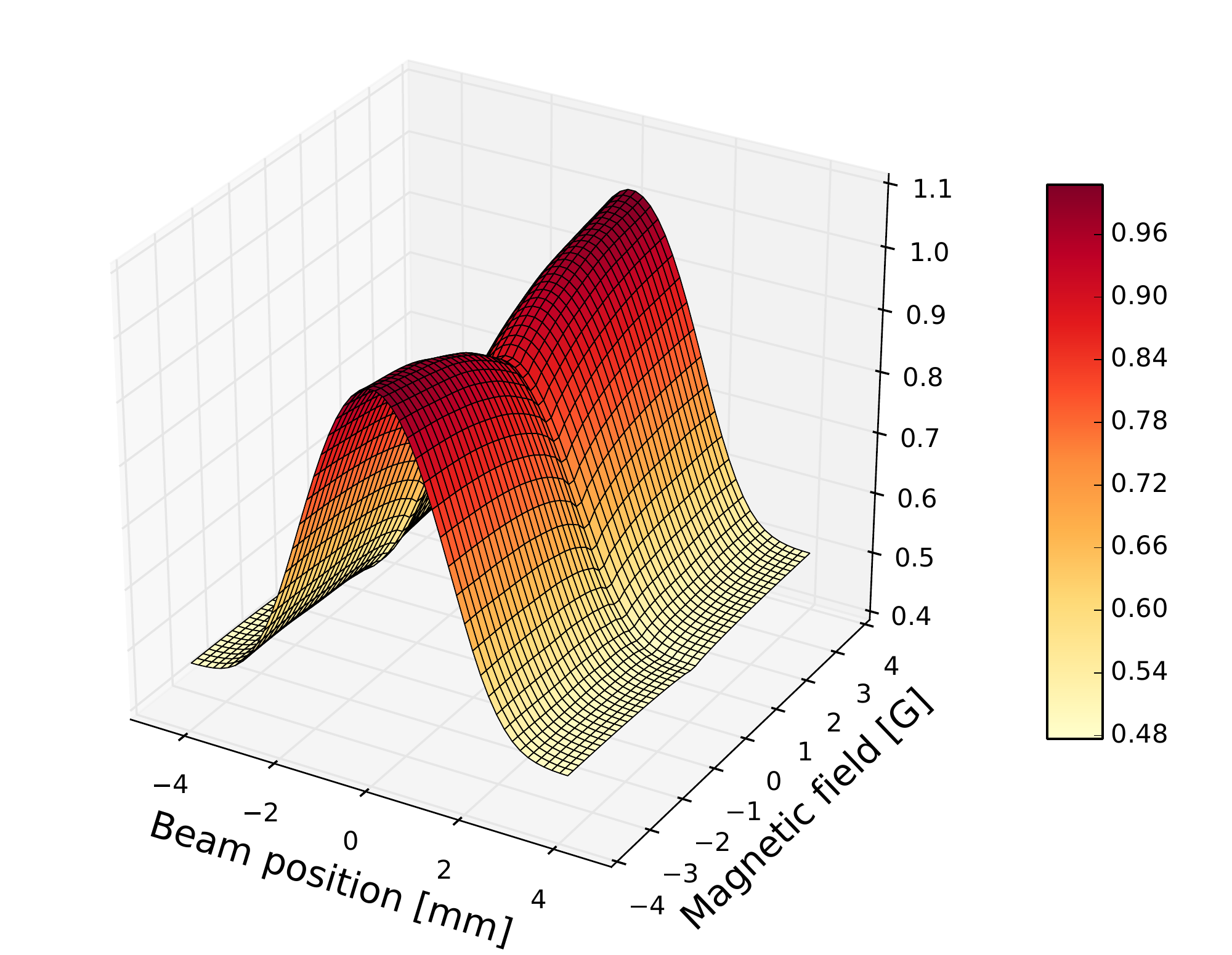}
	\caption{\label{fig:p44_s34} (Color online) Magneto-optical resonances produced for various positions of the probe laser beam 
	($F_g=3\longrightarrow F_e=4$ transition of the $D_1$ line of cesium) with respect to the pump laser beam 
	($F_g=4\longrightarrow F_e=4$ transition of the $D_2$ line of cesium): 
	(a) experimental results and (b) theoretical calculations.   }
\end{figure*}

In order to test our theoretical model of the spatial distribution of fluorescence from atoms in an intense, extended pumping beam, 
we decided to record magneto-optical resonances  
from various positions in the pumping beam. The experimental setup is shown in 
Fig.~\ref{fig:exp_setup}. To visualize these data, surface plots were generated where one horizontal 
axis represented the magnetic field and the other, the position of the probe beam relative to the pump beam axis. The 
height of the surface represented the fluorescence intensity. In essense, the surface consists of a series of 
magneto-optical resonances recorded for a series of positions of the probe beam axis relative the the pump beam axis. 
Fig.~\ref{fig:p44_s43} shows the results for experiments [(a)] and calculations [(b)] for which the pump beam was tuned to the 
$F_g=4\longrightarrow F_e=4$ transition of the Cs $D_2$ line and the probe beam was tuned to the
$F_g=4\longrightarrow F_e=3$ transition of the Cs $D_1$ line. 
%Experimental signals are shown in the first plot of each pair, and the results of calculations are shown in the second plot. 
One can see that the theoretical plot reproduces qualitatively all the features of the experimental measurement. 
Similar agreement can be observed when the  probe beam was tuned to the $F_g=3\longrightarrow F_e=4$ transition 
of the Cs $D_1$ line, as 
shown in Fig.~\ref{fig:p44_s34}. 
\vfill
\break

\section{\label{Conclusions:level1}Conclusions}
We have set out to model magneto-optical signals more accurately at laser intensities significantly higher than the saturation 
intensity by dividing the laser beam into concentric circular regions and solving the rate equations for Zeeman coherences in each region while 
taking into account the actual laser intensity in that region and the transport of atoms between regions. This approach was used to 
model magneto-optical resonances for the $F_g=2 \longrightarrow F_e=1$ transitions of the $D_1$ line of $^{87}$Rb, comparing the 
calculated curves to measured signals. 
We have demonstrated that good agreement between theory and experiment can be achieved up to Rabi frequencies of at least 25~MHz, 
which corresponds to a laser intensity of 200 mW/cm$^2$, or more than 40 times the saturation intensity of the transition.
As an additional check on the model, we have studied the spatial distribution of the fluorescence intensity within a laser beam theoretically 
and experimentally. The results indicated that at high laser power densities, the maximum fluorescence intensity is not produced 
in the center of the beam, because the atoms have been pumped free of absorbing levels prior to reaching the center. We compared
experimental and theoretical signals of magneto-optical resonance signals obtained by exciting cesium atoms with a 
narrow, weak probe beam tuned to the $D_1$ transition at various locations inside a region illuminated by an intense pump beam 
tuned to the $D_2$ transition and obtained good qualitative agreement.

\begin{acknowledgments}
We gratefully acknowledge support from the Latvian Science Council Grant Nr. 119/2012 and 
from the University of Latvia Academic Development Project Nr. AAP2015/B013.
\end{acknowledgments}

\bibliography{rubidium}% Produces the bibliography via BibTeX.

\end{document}